\documentclass{aastex62}

\newcommand\lya{$\mathrm{Lyman}\,\alpha$}
\graphicspath{{./}{figures/}}

\received{Nov, 2019}
\revised{Nov, 2019}
\accepted{\today}
\submitjournal{ApJ}

\shorttitle{High-energy SED of TRAPPIST-1}
\shortauthors{Wilson et al.}

\begin{document}

\title{The Mega-MUSCLES Spectral Energy Distribution Of TRAPPIST-1}

\correspondingauthor{David J. Wilson}
\email{djwilson394\@gmail.com}

\author[0000-0001-9667-9449]{David J. Wilson}
\affil{McDonald Observatory, University of Texas at Austin, Austin, TX 78712}

\author[0000-0001-8499-2892]{Cynthia S. Froning}
\affiliation{McDonald Observatory, University of Texas at Austin, Austin, TX 78712}

\author[0000-0002-7119-2543]{Girish M. Duvvuri}
\affil{Department of Astrophysical and Planetary Sciences, University of Colorado, Boulder, CO 80309, USA}

\author[0000-0002-1002-3674]{Kevin France}
\affil{Department of Astrophysical and Planetary Sciences, University of Colorado, Boulder, CO 80309, USA}
\affil{Laboratory for Atmospheric and Space Physics, University of Colorado, 600 UCB, Boulder, CO 80309}

\author[0000-0002-1176-3391]{Allison Youngblood}
\affil{Goddard Space Flight Center, Greenbelt, MD 20771}
\affil{Laboratory for Atmospheric and Space Physics, University of Colorado, 600 UCB, Boulder, CO 80309}

\author[0000-0002-5094-2245]{P.\ Christian Schneider}
\affil{Hamburger Sternwarte, Gojenbergsweg 112, 21029 Hamburg }

\author{Zachory Berta-Thompson},
\affil{Department of Astrophysical and Planetary Sciences, University of Colorado, Boulder, CO 80309, USA}

\author{Alexander Brown}
\affil{Department of Astrophysical and Planetary Sciences, University of Colorado, Boulder, CO 80309, USA}

\author{Andrea P. Buccino}
\affil{ Dpto. de Física, Facultad de Ciencias Exactas y Naturales (FCEN), Universidad de Buenos Aires (UBA), Buenos Aires, Argentina}

\author{Suzanne Hawley}
\affil{Astronomy Department, University of Washington, Seattle, WA 98195, USA}

\author{Jonathan Irwin}
\affil{Harvard-Smithsonian Center for Astrophysics, 60 Garden St., Cambridge, MA 02138, US}

\author{Lisa Kaltenegger}
\affil{Astronomy Department, Cornell University, Ithaca, NY 14853, USA}

\author{Adam Kowalski}
\affil{Department of Astrophysical and Planetary Sciences, University of Colorado, Boulder, CO 80309, USA}
\affil{Laboratory for Atmospheric and Space Physics, University of Colorado, 600 UCB, Boulder, CO 80309}Sa
\affil{National Solar Observatory, University of Colorado at Boulder, 3665 Discovery Drive, Boulder, CO 80303}

\author{Jeffrey Linsky}
\affil{JILA, University of Colorado and NIST, Boulder, CO 80309-0440 USA}

\author{R.~O. Parke Loyd}
\affil{School of Earth and Space Exploration, Arizona State University, Tempe, AZ 85287}

\author{Yamila Miguel}
\affil{Leiden Observatory, P.O.\ Box 9500, 2300 RA Leiden, The Netherlands}


\author{J. Sebastian Pineda}
\affil{Department of Astrophysical and Planetary Sciences, University of Colorado, Boulder, CO 80309, USA}

\author{Seth Redfield}
\affil{Wesleyan University, Department of Astronomy and Van Vleck Observatory, 96 Foss Hill Dr., Middletown, CT 06459, USA}

\author{Aki Roberge}
\affil{Goddard Space Flight Center, Greenbelt, MD 20771}

\author{Sarah Rugheimer}
\affil{University of Oxford, Clarendon Laboratory, AOPP, Sherrington Road, Oxford, OX1 3PU, UK}

\author{Feng Tian}
\affil{State Key Laboratory of Lunar and Planetary Sciences, Macau University of Science and Technology, Macau, China}

\author{Mariela Vieytes}
\affil{ Instituto de Astronomía y Física del Espacio (CONICET-UBA), Buenos Aires, Argentina.}




\begin{abstract}
We present a 5\,\AA--100\,$\mu$m Spectral Energy Distribution (SED) of the ultracool dwarf star TRAPPIST-1, obtained as part of the Mega-MUSCLES Treasury Survey. The SED combines ultraviolet and blue-optical spectroscopy obtained with the \textit{Hubble Space Telescope}, X-ray spectroscopy obtained with \textit{XMM-Newton}, and models of the stellar photosphere, chromosphere, transition region and corona. A new Differential Emission Measure model of the unobserved extreme-ultraviolet spectrum is provided, improving on the Lyman $\alpha$--EUV relations often used to estimate the 100--911\,\AA\ flux from low-mass stars. We describe the observations and models used, as well as the recipe for combining them into an SED. We also provide a semi-empirical, noise-free model of the stellar ultraviolet spectrum based on our observations for use in atmospheric modelling of the TRAPPIST-1 planets.

\end{abstract}



\section{Introduction}
\label{sec:intro}

\begin{figure}
    \centering
    \includegraphics[width=\columnwidth]{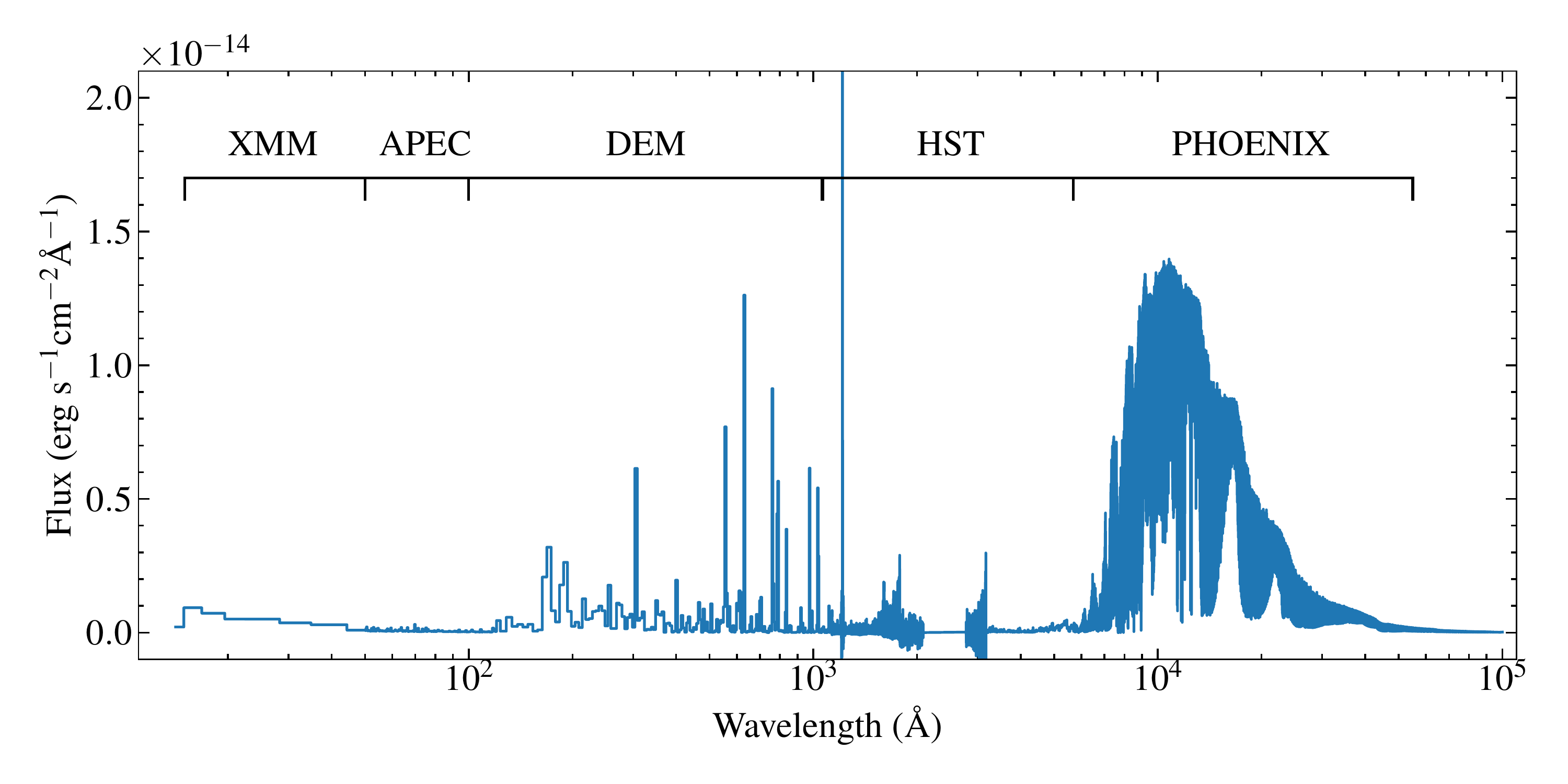}
    \caption{SED of TRAPPIST-1 with all data and models at native resolutions. The sources for each section of the spectrum are labeled above it.}
    \label{fig:t1_sed}
\end{figure}{}

Among the thousands of planetary systems that have been discovered over the past two and a half decades, TRAPPIST-1 is a standout case. Discovered by the  
TRansiting Planets and PlanestIsimals Small Telescope \hbox{(TRAPPIST)} survey in 2015 \citep{gillonetal16-1}, the system is comprised of an M8 ultracool dwarf star orbited by seven planets, all of which have similar masses and radii to Earth and Venus \citep{gillonetal17-1, wangetal17-1}. The planets are almost exactly coplanar, have orbital periods ranging between 1.5 and 18.8 days, and are all in an orbital resonance with at least one other planet \citep{lugeretal17-1}. At a distance of $12.43\pm0.02$\,pc and $r_{\mathrm{mag}}=17.87\pm0.01$ \citep{chambersetal16-1}, the system presents a challenging but achievable target for transit spectroscopy observations of the planets, both now with the Hubble Space Telescope (HST) \citep{dewitetal16-1} and in the future with JWST \citep{barstow+irwin16-1, morleyetal17-1}. Three or four of the planets orbit at distances where the energy received from the star is such that liquid water might persist on their surfaces. The TRAPPIST-1 system therefore offers opportunities for comparative planetology to test models of planetary habitability, biosignatures and even, given the small orbital separations between the planets, transfer of material and/or life between the planets \citep{verasetal18-1}.

A complete evaluation of the potential habitability of the TRAPPIST-1 planets requires comprehensive knowledge of the parent star. This has proven challenging, with uncertainties remaining over, for example, the star's age \citep{burgasseretal17-1, gonzalesetal19-1}, activity \citep{vidaetal17-1},  and rotation period \citep{roettenbacheretal17-1}. Of particular importance, given the close proximity of the planets to the star, is the stellar magnetic activity and the resulting X-ray and ultraviolet emission. High-energy radiation can influence the retention and chemistry of planetary atmospheres as well as surface survival conditions \citep{rugheimeretal15-1, migueletal15-1, omalleyjamesetal17-1}. However, TRAPPIST-1 is extremely faint at short wavelengths, making detailed characterisation of the high-energy environment in the system challenging. \citet{wheatleyetal17-1} observed TRAPPIST-1 with \textit{XMM-Newton} (\textit{XMM}), finding variable X-ray luminosity with intensity similar to the modern quiescent Sun. Because of their proximity to the host stars, the planets would therefore experience XUV intensities much higher than the Earth, sufficient to significantly alter their atmospheres and strip away hydrogen from water in their atmospheres and (if present) oceans \citep{ribasetal16-1, airapetianetal17-1}. \citet{bourrieretal17-1, bourrieretal17-2} obtained time series observations of the 1215.67\AA\ \lya\ hydrogen emission line with \textit{HST}, finding that it evolved over a three-month timescale but with no evidence for hydrogen (and thus water) escape from TRAPPIST-1c, which transited during their observations. \citet{peacocketal19-1} used the PHOENIX stellar atmosphere code to model the chromosphere and transition region of TRAPPIST-1, scaling it to the \citet{bourrieretal17-1} \lya\ measurement and to distance-adjusted {\em GALEX} observations of stars with a similar spectral type. They found that the flux emitted between 100--912\,\AA\ varies by an order of magnitude depending on which calibrator was used. These studies demonstrate that accurately accounting for the effects of high-energy radiation on the TRAPPIST-1 planets requires spectroscopic observations at all accessible wavelengths. 

Mega-MUSCLES (Measurements of the Ultraviolet Spectral Characteristics of Low-Mass Exoplanetary Systems) is an \textit{HST} Treasury program obtaining 5\,\AA--100\,$\mu$m spectral energy distributions (SEDs) of a representative sample of 12 M\,dwarfs, covering a wide range of stellar mass, age, and planetary system architecture and extending the original 11-star MUSCLES program \citep{franceetal16-1,youngbloodetal16-1, loydetal16-1} to stars with lower masses, higher activity and/or faster rotation rates. Here we present the Mega-MUSCLES SED of TRAPPIST-1 (Figure \ref{fig:t1_sed}), comprised of ultraviolet and X-ray spectroscopy with \textit{HST} and \textit{XMM}, along with state-of-the-art model spectra. We discuss the changes made to the data processing and stellar emission modelling implemented for Mega-MUSCLES compared with MUSCLES, present a semi-empirical model for use in model atmosphere simulations, and compare the observed SED to the \citet{peacocketal19-1} models.

\section{Observations}
\label{sec:obs}

\begin{table*}
\centering
\caption{Summary of observations from. Dataset numbers are given for retrieval from MAST (\url{https://archive.stsci.edu/hst/}) or the XMM-Newton Science Archive (\url{http://nxsa.esac.esa.int/nxsa-web/\#home}) for HST and XMM respectively.}  
\begin{tabular}{lcccccc}\\
\hline
Date & Instrument & Grating &  Central Wavelength (\AA) & Start Time (UT) & Total Exposure Time (s) & Dataset \\
\hline 
\textit{HST} & & & & & & \\
2017-12-15 & COS  &	G230L &	2950.000 & 07:19:18 & 2731  & LDLM42010 \\
2018-12-08 & COS  &	G160M &	1577.000 & 06:53:42 & 1658  & LDLM39010 \\
2018-12-08 & COS  &	G160M &	1611.000 & 08:07:13	& 5352  & LDLM39020 \\
2018-12-09 & STIS &	G430L &	4300.000 & 01:49:27 & 1795  & ODLM41010 \\
2018-12-09 & STIS &	G140M &	1222.000 & 04:46:12	& 2710  & ODLM41030 \\
2018-12-09 & STIS &	G140M &	1222.000 & 06:21:32 & 2710  & ODLM41040 \\
2018-12-09 & STIS &	G140M &	1222.000 & 07:56:53 & 2710  & ODLM41050 \\
2018-12-10 & COS  &	G130M &	1291.000 & 03:22:34 & 12403 & LDLM40010 \\
2019-06-07 & COS  &	G160M &	1577.000 & 19:54:25 & 1598  & LDLMZ6010 \\
\textit{XMM} & & & & & & \\
2018-12-10 & EPIC & -- & -- &  10:42:57 &  24600 & 0810210101 \\
	
\hline

\hline
\end{tabular}
\label{tab:hst_obs}
\end{table*}

We observed TRAPPIST-1 with the Cosmic Origins Spectrograph \citep[COS,][]{greenetal12-1} and the Space Telescope Imaging Spectrograph \citep[STIS,][]{woodgateetal98-1} onboard the \textit{Hubble Space Telescope} (\textit{HST}) on 2017~December~15, 2018~December~08--12 and 2019~June--08, for a total exposure time of 36379\,s. The COS gratings used were G160M (8608\,s), G130M (12404\,s) and G230L (2731\,s), and the STIS gratings were G430L (1795\,s) and G140M (10841\,s). Combined, these spectra cover the wavelength range 1130--5700\,\AA\, except for a gap between 2080--2790\,\AA\ which is not covered by the COS NUV detector and is too faint for STIS NUV observations. The HST observations are summarised in Table \ref{tab:hst_obs}. With the exception of the STIS/G430L exposure, the observations were obtained using photon-counting detectors in TIME-TAG mode. We extracted light curves from each spectrum to search for and potentially remove contributions from flares or other stellar activity, but found no significant variation.

For the COS G160M and G130M observations, variations in target position in the aperture between each orbit induce slight differences in wavelength calibration. To remove this effect we cross-correlated known emission lines in each x1d spectrum to shift each spectrum onto a single wavelength scale before coadding. Doing so provides a small increase in S/N and resolution compared with the x1dsum files produced by  the {\sc CalCOS} pipeline. 

Four spectra were obtained with the STIS G140M grating, covering the Lyman\,$\alpha$ hydrogen line with a spectral range of 1195--1249\,\AA\ and R$\sim$10000. The automated reduction pipeline failed to identify the spectral trace, so we used the flt images to visually identify the spectrum in each exposure, before re-extracting the spectrum with the {\sc stistools} x1d routine, fixing the ``a2center'' keyword to the identified spectrum position. In one spectrum the trace could not be visually identified. Lyman \,$\alpha$ line fluxes in the spectra obtained immediately before and after the non-detection were similar, so the non-detection is unlikely to be due to intrinsic variability and probably an instrumental effect such as inaccurate slit positioning. The non-detection was therefore discarded, and the remaining three spectra coadded into the final G140M spectrum used hereafter.

TRAPPIST-1 has been observed multiple additional times with the STIS G140M grating both prior to and since our observations \citep{bourrieretal17-1, bourrieretal17-2}. We initially intended to combine all of the available spectra, but we found the detected Lyman\,$\alpha$ flux to be variable, with the final coadded spectrum highly dependent on which subspectra were chosen for inclusion. Epoch-to-epoch changes are somewhat beyond the scope of this paper and we do not wish to preempt the teams leading the ongoing G140M observations, so we decided to use only the data obtained as part of the Mega-MUSCLES program. This has the advantage of ensuring that all of the data in the SED is contemporaneous, at the cost of improved S/N for the Lyman\,$\alpha$ measurement. We will reevaluate this decision when a full analysis of the G140M data is available and update the Mega-MUSCLES High-Level Science Products accordingly. 

We further observed TRAPPIST-1 with {\em XMM-Newton} (\textit{XMM}) using the EPIC instrument with thin filters for 23\,ks on 2018~December~10, overlapping in time with the COS G130M observations. TRAPPIST-1 was detected with an average count rate of 3\,counts	\,ks$^{-1}$, less than the average count rate of $\approx 10-20$\,counts	\,ks$^{-1}$ found by \citet{wheatleyetal17-1}. Analysis of all available archival data suggested that the \citet{wheatleyetal17-1} result was dominated by a strong flare(s), thus we take our observation to represent the ``typical'' or quiescent X-ray flux.

\begin{figure}
    \centering
    \includegraphics[width=0.5\columnwidth]{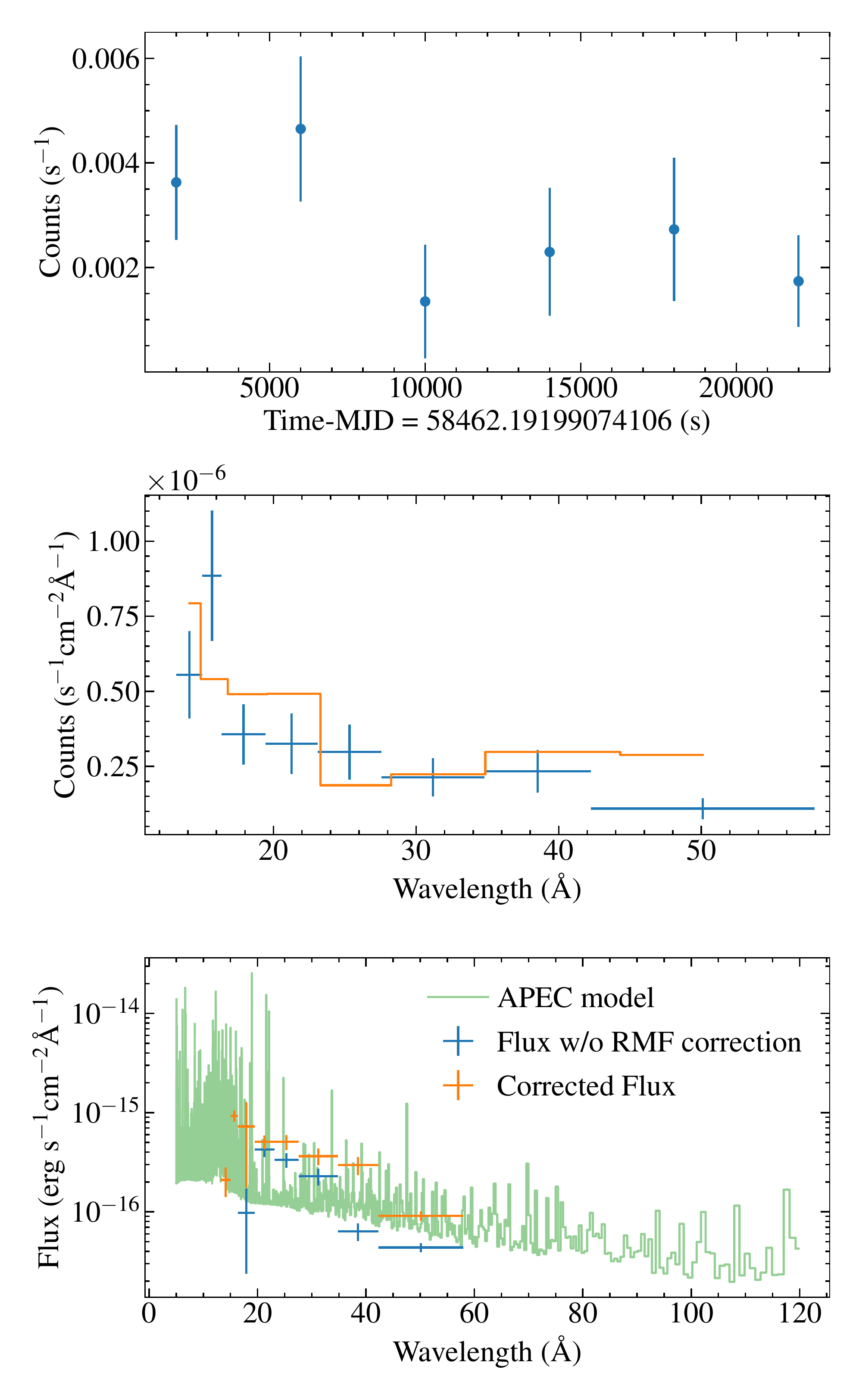}
    \caption{Top: X-ray light curve of TRAPPIST-1 observed on 2018~December~10. Middle: \textit{XMM} spectrum of TRAPPIST-1 (blue) compared with the APEC model (orange). The model has been binned into the same bins as the spectrum. Bottom: Full APEC model used to correct the x-ray spectrum and provide the 50--120\,\AA\ range of the SED, along with the raw and RMF-corrected flux conversions described in the text.}
    \label{fig:xmm}
\end{figure}

\section{SED}
Figure \ref{fig:t1_sed} shows the full spectral energy distribution (SED) and the wavelength ranges covered by the different observational and model sources. The methods used to produce the SED for TRAPPIST-1 and the rest of the Mega-MUSCLES sample are based on those used for the MUSCLES survey \citep{loydetal16-1}, but with multiple differences reflecting developments in modeling techniques since MUSCLES and, in the case of TRAPPIST-1, the faintness of the target in the ultraviolet. Below we  describe in full the production of the TRAPPIST-1 SED. When discussing specific wavelength regions we follow the definitions used for the MUSCLES program by \citet{franceetal16-1}: X-ray: 5--100\,\AA; EUV: 100--911\,\AA; FUV: 911--1700\,\AA; NUV: 1700--3200\,\AA\ and optical/IR: 3200\,\AA--100.0\,$\mu$m.

\subsection{Variability}
Late-type M\,dwarfs are notably active stars, with regular flares and high photometric variability \citep{paudeletal18-1}. Even stars that appear photometrically quiet in optical surveys have been shown to be active in the ultraviolet \citep{loydetal18-1, franceetal20-1}. \cite{vidaetal17-1} identified 42 white-light flaring events in the \textit{K2} light curve of TRAPPIST-1, approximately 0.5\,d$^{-1}$, although \cite{gillonetal17-1} only detected two flares in 20 days of infrared \textit{Spitzer} observations taken at a different epoch, indicating that the flare rate is time and/or wavelength dependent. We must therefore consider activity when assessing the validity of the SED presented here. We detected no flares in any of our ultraviolet and x-ray observations and thus consider the SED to represent TRAPPIST-1 in a ``quiescent'', non-flaring state. Flares have been detected in previous x-ray observations by \cite{wheatleyetal17-1}, and potentially in broadband NUV \textit{Swift} photometry by \cite{beckeretal20-1}. Determining how representative the ultraviolet spectra presented here are of the the time-averaged emission of TRAPPIST-1 will require sustained monitoring at multiple wavelengths, a significant challenge considering the limitations of current ultraviolet observatories. Therefore, although we are confident that the Mega-MUSCLES SED is an accurate measurement of the quiescent emission of TRAPPIST-1, efforts to model the atmospheres of the TRAPPIST-1 planets should acknowledge that we have poor constraints on how that emission changes during a flare, and how often such flares occur.

\subsection{X-ray}
The \textit{XMM} spectrum was fit using the {\tt{XSPEC}} v12.10 package \citep{arnaud96-1} with models generated using the Astrophysical Plasma Emission Code v3.0.9 \citep[APEC,][]{smithetal01-1, fosteretal12-1}, with the data binned in variable energy bin widths such that each bin contained the same number of counts. The data were fit using a two-temperature model ($k$T = 0.2, 0.4) and a metallicity of 0.4 Solar. Due to the low number of photons, the only free parameters used were the normalisation of the two temperature components. We used the standard {\tt{XSPEC}} background subtractions, although we found that ignoring the background changed the derived flux by less than 0.1 dex. 

Ordinarily, the data could be converted from counts $N$ into physical units $F$ by dividing by the the effective area $A$ of the instrument as a function of energy $E$, i.e.:
\begin{equation}
F(E) \propto \frac{N(E)}{A(e)\,T_{\mathrm{exp}}} 
\label{eq:ea}    
\end{equation}

However, for cool spectra such as TRAPPIST-1, issues arise at the low energy boundary ($0.2$\,keV), because more photons are discarded when detected below the energy cutoff than photons being recorded at higher than nominal energies due to the highly asymmetric energy response matrix (RMF). A priori, we do not know the magnitude of this effect, but with a reasonable model spectrum, we can estimate the energy-dependent fraction of photons being lost to this effect. We use this (mildly) model-dependent knowledge to correct Equation \ref{eq:ea} to: 

\begin{equation}
F(E) \propto \frac{N(E)}{A(e)\,T_{\mathrm{exp}}\,C(E)} 
\label{eq:eac}    
\end{equation}

Where the correction $C$ captures the asymmetry of the spectrum and the RMF. For a symmetric RMF and a flat spectrum, $C(E)$ would be 1.0. Consider a spectrum consisting of single emission line at 0.3\,keV and low energy cutoff for the spectrum of 0.2\,keV: The recorded photon rate will be lower than expected, because a sizable fraction of the photons will have reconstructed energies below 0.2\,keV and will thus be missing from the spectrum. This is not an effective area but an RMF effect. In fact, the shape of the XMM-Newton RMF at low energies (0.1--0.3\,keV) is such that more photons will be detected below their nominal energy than above, so that even for a flat spectrum, one would need to correct for the RMF effect. In essence, this procedure ensures that the corrected spectrum reproduces the model flux, which would not be the case when using Equation \ref{eq:ea}. However, we caution that this ``corrected'' spectrum still includes RMF effects and is meant for illustrative purposes in physical units while the best-fit model spectrum captures the emitted spectrum much better.

\subsection{Extreme-Ultraviolet and DEM modelling}
The most significant departure from the MUSCLES procedure is in the Extreme-Ultraviolet (EUV), where we have replaced the \citet{linskyetal14-1} empirical scaling relations with a Differential Emission Measurement (DEM) model, which estimates the chromospheric, transition region and coronal emission based on the strength of the detected lines in the FUV spectrum and the X-ray flux. A model spectrum is required as the region spanning 120--1100\,\AA\ is unobservable, both physically due to absorption from interstellar hydrogen between 400-900\,\AA, and a lack of currently operating instruments that can observe the ranges 120--400\,\AA\ and 900--1100\,\AA\ (\textit{HST}/COS and \textit{Chandra} do have some modes that extend down to 900\,\AA\ and up to 175\,\AA\ respectively, but these were not sensitive enough to be practical for this program). The DEM model is described in detail in Section \ref{sec:dem}. 

\subsection{Far- and Near- Ultraviolet}

\begin{figure}
    \centering
    \includegraphics[width=\columnwidth]{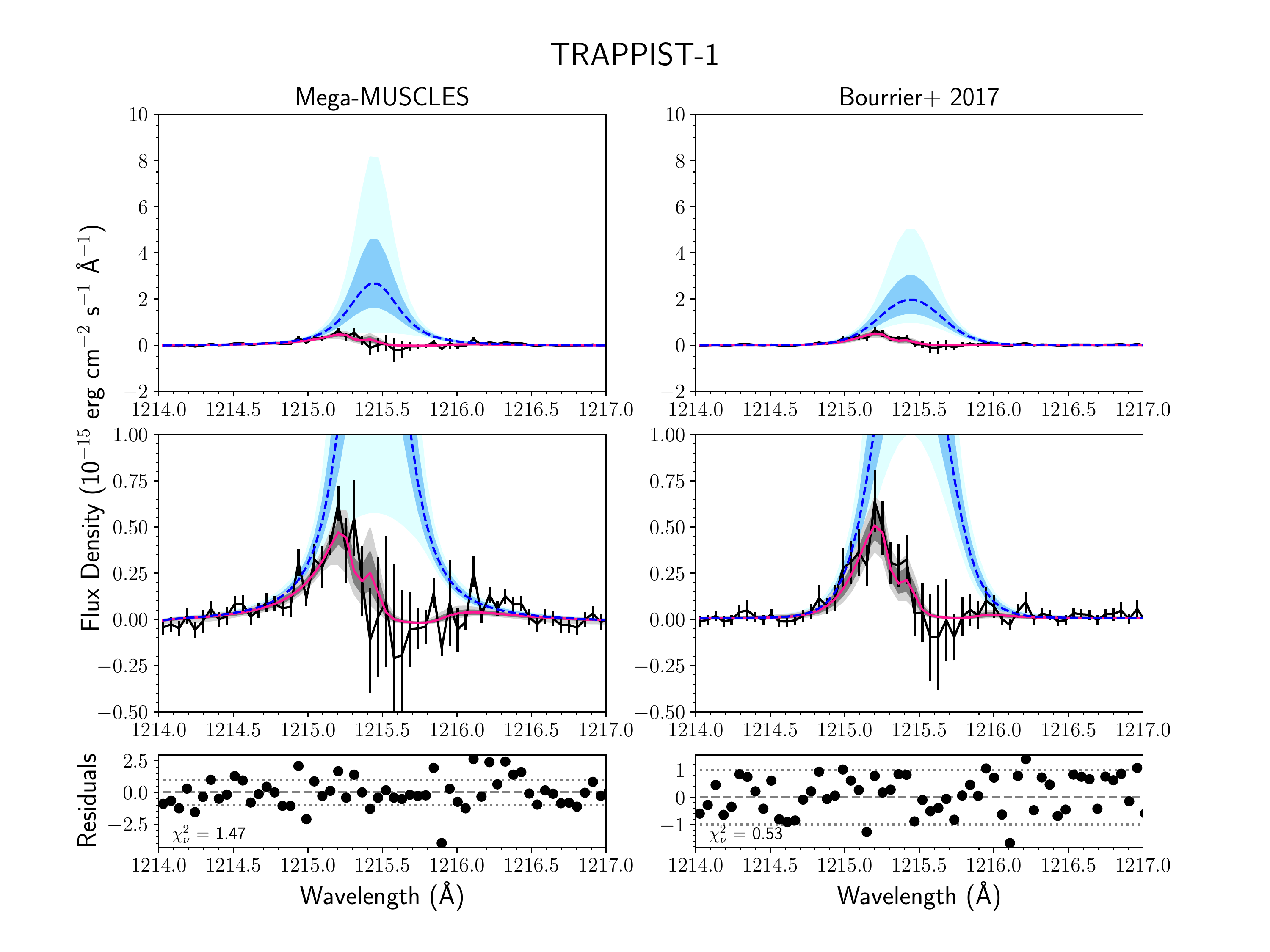}
    \caption{Best-fit models and intrinsic Lyman alpha profiles are shown from a simultaneous fit to the new spectrum presented in this work and the spectrum presented in \cite{bourrieretal17-1}. The ISM properties were forced to be the same for the two spectra, but the intrinsic emission profiles were allowed to vary, to account for either intrinsic stellar variability or differences in the flux calibration between the two datasets. The solid pink lines show the best fits to the data (black lines with error bars), with the dark and light shaded regions corresponding to the 68\% and 95\% confidence intervals, respectively. The dashed blue lines show the intrinsic emission profiles corresponding to the best fit models, and the dark blue and light blue shaded regions represent the 68\% and 95\% confidence intervals, respectively. The bottom panels show the residuals of the fit, or the best fit models subtracted from the data and divided by the data uncertainty. The reduced chi-squared values are printed in the same panels.}
    \label{fig:lya}
\end{figure}

\subsubsection{Lyman $\alpha$ reconstruction}
The \ion{H}{1}\,1215.67\,\AA\ Lyman $\alpha$ line is clearly visible in our coadded STIS G140M spectrum, but is heavily affected by both interstellar absorption and terrestrial airglow. We reconstructed the full Lyman $\alpha$ profile with a technique derived from \citet{youngbloodetal16-1}, where we simultaneously fit a model of the interstellar absorption and a model of the intrinsic stellar emission with a Markov Chain Monte Carlo method (\texttt{emcee}; \citealt{foreman-mackeyetal13-1}). In order to benefit from the increased S/N offered by spectra taken before the Mega-MUSCLES observation while simultaneously accounting for possible intrinsic stellar variability, we fit both the new Mega-MUSCLES observation and the observation from \cite{bourrieretal17-1} simultaneously. We use a Voigt profile for the stellar emission as well as for the ISM absorption, but require the ISM absorption to be the same for both observations. To account for wavelength solution errors between the two observations, we do not fix the radial velocities to the same value, but rather require the radial velocity offset between the emission and ISM absorption to be the same between the two observations. Based on the measured stellar radial velocity -56.3$\pm$1 km s$^{-1}$ \citep{reiners+basri09-2} and the predicted ISM radial velocity along TRAPPIST-1's sightline (-1.25$\pm$1.37 km s$^{-1}$, \citealt{redfield+linksy08-1}), we give the radial velocity offset parameter a Gaussian prior with mean +55.05$\pm$1.70 km s$^{-1}$, obtained by adding the stellar and ISM radial velocities and adding their uncertainties in quadrature. We fix the Doppler b value of the ISM absorption profile to 11.5\,km\,s$^{-1}$ and D/H$=1.5\times10^{-5}$, both standard values for the local ISM \citep{wood04-1}. All other parameters were varied with uniform priors. 

Figure \ref{fig:lya} shows the spectrum and reconstructed line profile. We report the median and the 68\,per\,cent confidence interval as our best fit values and $1\sigma$~error bars. We found an integrated flux of $F_{\mathrm{Ly\alpha}} = (1.40^{+0.60}_{-0.36})\times10^{-14}\,\mathrm{erg}\,\mathrm{s}^{-1}\,\mathrm{cm}^{-2} $ for the Mega-MUSCLES observation, and $F_{\mathrm{Ly\alpha}} = (1.09^{+0.40}_{-0.27})\times10^{-14}\,\mathrm{erg}\,\mathrm{s}^{-1}\,\mathrm{cm}^{-2} $ for the \cite{bourrieretal17-1} observation. This value is larger but within $1\,\sigma$ of the flux ($(8.18^{+1.64}_{-3.27})\times10^{-15}\,\mathrm{erg}\,\mathrm{s}^{-1}\,\mathrm{cm}^{-2} $) found by \citet{bourrieretal17-1}, and is explained by our fit's larger column density (log$_{10}$ N(HI) = 18.4$\pm$0.1, compared to 18.3$\pm$0.2 from \citealt{bourrieretal17-1}). However, these column densities and reconstructed fluxes are consistent with the result from \citealt{bourrieretal17-2} within 1-sigma, and we do not detect any statistically significant intrinsic stellar variability between the two observations. 

\subsubsection{COS spectra}
The COS FUV spectrum was contaminated by airglow from Lyman $\alpha$ and \ion{O}{1} over the wavelength ranges 1214--1217\,\AA\ and 1301--1307\,\AA\ respectively. Both ranges were removed and replaced by the reconstructed Lyman $\alpha$ profile in the first case and by a polynomial fit to the spectrum on either side in the second. As we therefore have no information at these wavelengths, for the rest of this paper we assume that there is zero flux from \ion{O}{1} at TRAPPIST-1.       

The COS NUV observations covered the wavelength ranges 1700--2100\,\AA\ and	2800--3200\,\AA\ leaving a 700\,\AA\ gap. This gap is partially covered by a second order spectrum spanning 1950--2150\,\AA, but the signal was so weak that we chose not to include it. The gap was filled with a polynomial fit to the two wavelength regions, with the range  2790--2805\,\AA\ masked out to remove contributions from the \ion{Mg}{2}\,2800\,\AA\ lines. 

\subsection{Optical to infrared}

\begin{figure}
    \centering
    \includegraphics[width=\columnwidth]{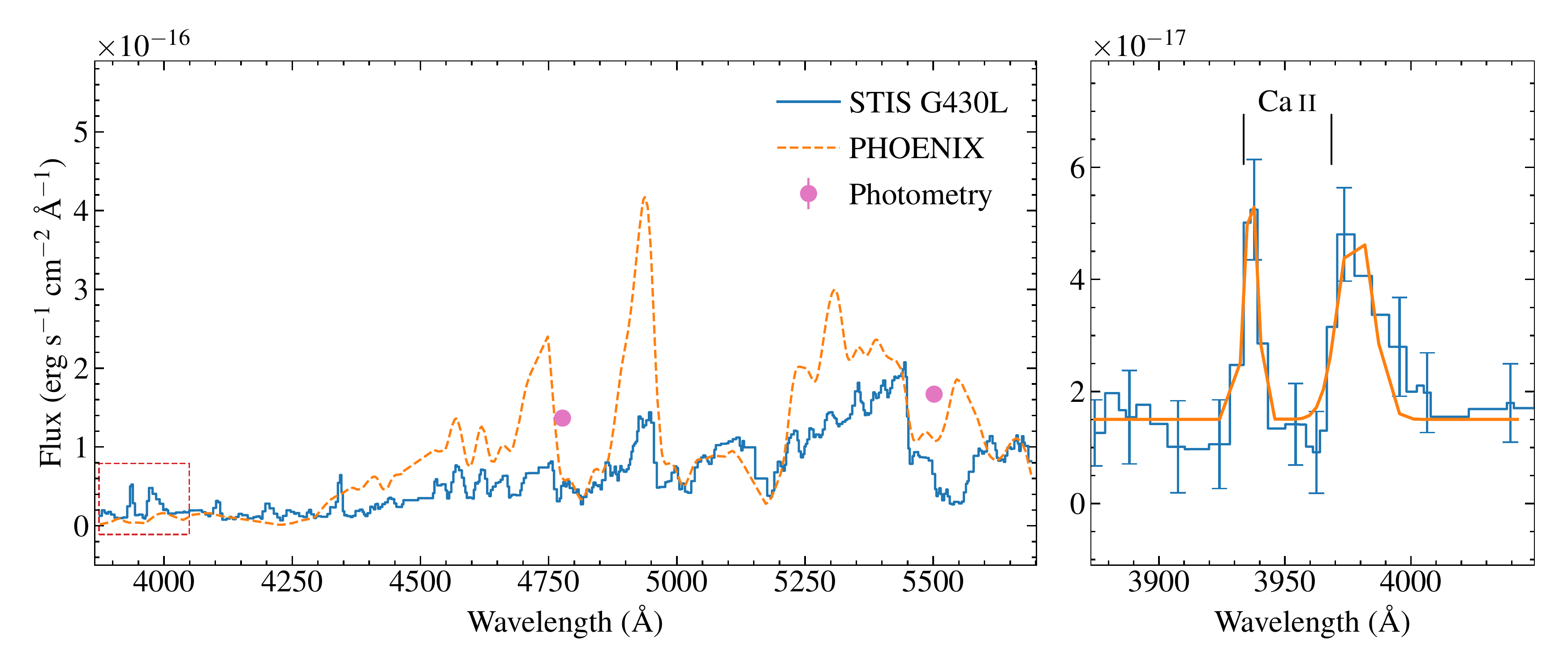}
    \caption{Left: STIS G140L spectrum (blue) compared with  Pan-STARSS \textit{g} and \textit{Gaia} $B_{p}$ photometry \citep{gonzalesetal19-1} and the PHOENIX model (orange). The spectrum has been smoothed by a factor 2 for clarity, and the PHOENIX model been convolved to the resolution of the observed spectrum. Right: Enlarged view of the region in the red box showing \ion{Ca}{2} H\&K  emission lines, along with the Gaussian fits (orange) used to measure the emission line fluxes (Table \ref{tab:t1_line_fluxes}). }
    \label{fig:g430l}
\end{figure}

\begin{figure}
    \centering
    \includegraphics[width=\columnwidth]{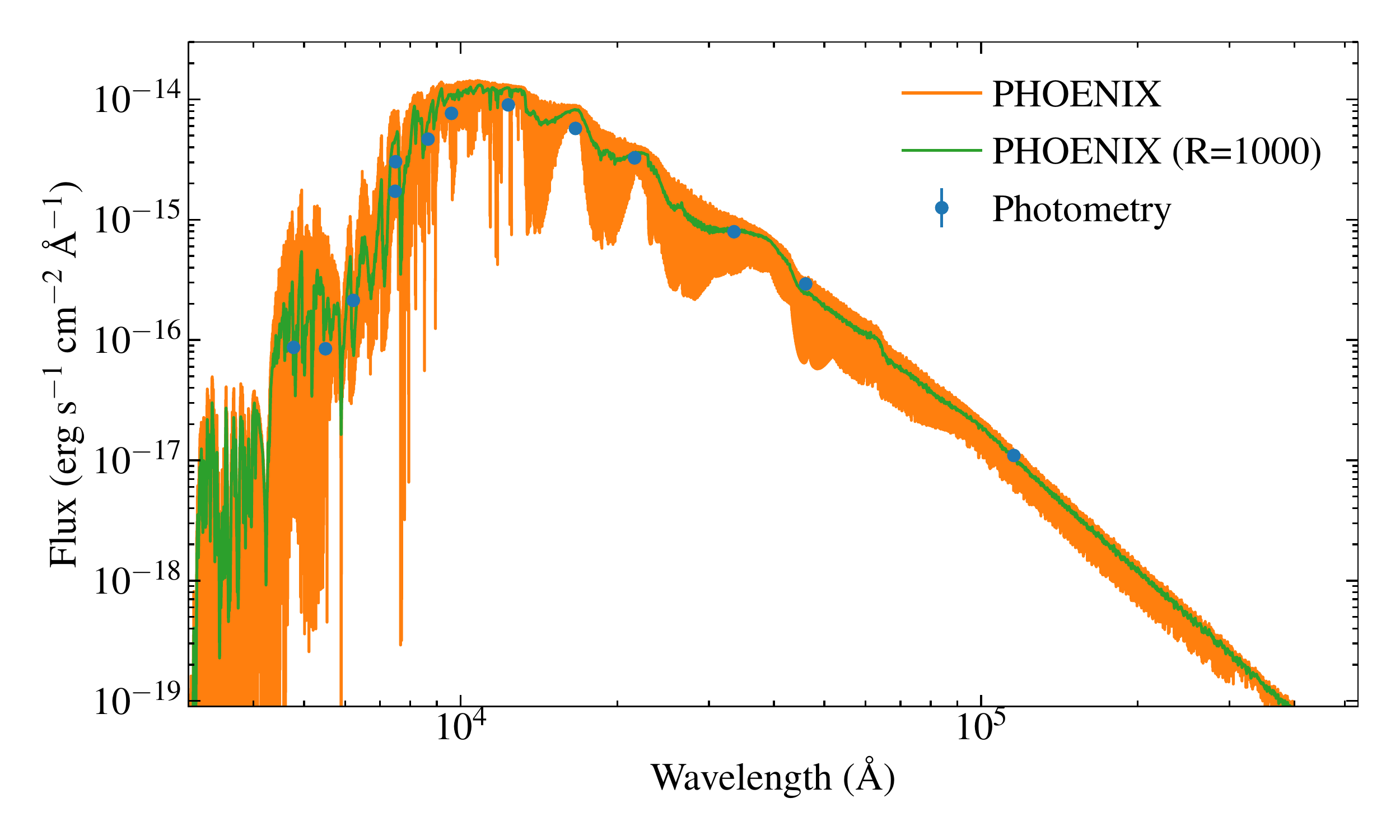}
    \caption{PHOENIX model compared with the photometry from Table 1 of \citet{gonzalesetal19-1}. The model in green has been convolved to a resolution of R=1000 for easier comparison with the photometry.}
    \label{fig:phx}
\end{figure}

\subsubsection{STIS G430L}
Figure \ref{fig:g430l} shows the STIS G430L spectrum of TRAPPIST-1. The spectrum was obtained with a single exposure on a CCD detector, and thus has a slightly different treatment to the rest of the HST data obtained with photon-counting detectors. Although the spectrum in principle covers the wavelength range 2900--5700\,\AA, in practice the combination of low detector throughput and the decreasing flux from the target at short wavelengths results in an effective non-detection of the blue end of this range, with almost half of the pixel bins containing negative flux values. We therefore removed all points where the mean flux/flux error ratio of the 30 surrounding bins was less than one. This results in a cutoff at 3872\,\AA, similar to the fixed 3850\,\AA\ cutoff used by \cite{loydetal16-1} for the MUSCLES SEDS.       

\cite{loydetal16-1} also found that the STIS G430L spectra obtained for the MUSCLES program had systematically lower integrated fluxes than photometric measurements of their respective stars, which they suggested was due to imperfect alignment on the STIS slit. In an attempt to correct this, the Mega-MUSCLES G430L spectra were obtained with the slit width set to 0.2", twice that used for MUSCLES. To validate that the fluxes of the Mega-MUSCLES G430L spectra are correct, we have obtained extensive flux-calibrated spectroscopy with various ground-based instruments, which are in good agreement with Mega-MUSCLES spectra for the same stars (Mega-MUSCLES Collaboration in prep.). However no comparison spectra are available for TRAPPIST-1 itself, so we have additionally compared the G430L spectra with available photometry. Of the filters for which photometric data is available, the $g$ and $B$ filters have band passes that fall nearly entirely in the G430L range. We retrieved $B$, SDSS\,$g$ and PanSTARS\,$g$ magnitudes from Vizier and compared them with synthetic photometry calculated by integrating the spectrum over the respective band passes. We found that the spectrum underpredicted the photometry by ratios of $0.5\pm0.1$, $0.85\pm0.02$ and $0.6\pm0.01$ for $B$, SDSS\,$g$ and PanSTARS\,$g$ respectively. Given that the differences are small and that photometry at short optical wavelengths is more likely to be affected by stellar activity than at longer wavelengths \citep{paudeletal18-1} we chose not to scale the spectra based on the photometry. An additional test is provided by the PHOENIX model, described in more detail below. As Figure \ref{fig:g430l} shows, the PHOENIX model is in good agreement with the TRAPPIST-1 data at regions of relatively low flux, but over-predicts the regions of higher flux. We speculate that this is due to an incomplete treatment of opacity at these wavelengths in the model. \citet{lanconetal20-1} reported similar discrepancies between PHOENIX models and blue spectra of cool stars. For our purposes here the disagreement cannot be fixed by scaling the spectrum, so we leave the flux calibration as is.

The G430L spectrum also contains features consistent with emission from the \ion{Ca}{2} H\&K lines at 3968.4673 and 3933.6614\,\AA. The Ca\,H line is blended with emission from the H\,$\epsilon$ line so is given as an upper limit in Table \ref{tab:t1_line_fluxes}.

\subsubsection{PHOENIX}

Wavelengths from 5700\AA{} to 100\,$\mu$m are filled with a PHOENIX photospheric model spectrum from the Lyon BT-Settl [DIR]	CIFIST2011\_2015 grid\footnote{\url{https://phoenix.ens-lyon.fr/Grids/BT-Settl/}}\citep{allard16-1, baraffeetal15-1} as no ground-based spectroscopy of TRAPPIST-1 contemporaneous with our \textit{HST} and \textit{XMM} observations is available. 
\citet{gonzalesetal19-1} used distance-calibrated spectra and photometry to obtain atmospheric parameters of TRAPPIST-1 of $T_{\mathrm{eff}} = 2628 \pm 42$\,K and $\log g = 5.21\pm0.06$\,dex. We obtained the four closest models ($T_{\mathrm{eff}}, \log g = $ 2600, 5.0; 2700, 5.0; 2600, 5.5; 2700, 5.5) and linearly interpolated them onto the measured parameters using the {\sc scipy} griddata routine. The model flux is then scaled by the square of the ratio of the measured radius and distance of TRAPPIST-1 \citep[$1.16\pm0.03\,R_{\mathrm{Jup}}$ and $12.43\pm0.02$\,pc respectively, ][]{gonzalesetal19-1}. The BT-Settl models extend to 1000\,$\mu$m, but to avoid the SED file size becoming too large we truncated the model at 100\,$\mu$m. The removed flux contributed less than one per\,cent of the total integrated flux of the model. Figure \ref{fig:phx} compares the PHOENIX  model with the data from \citet{gonzalesetal19-1} used to measure the atmospheric parameters.

\begin{figure}
    \centering
    \includegraphics[width=0.65\columnwidth]{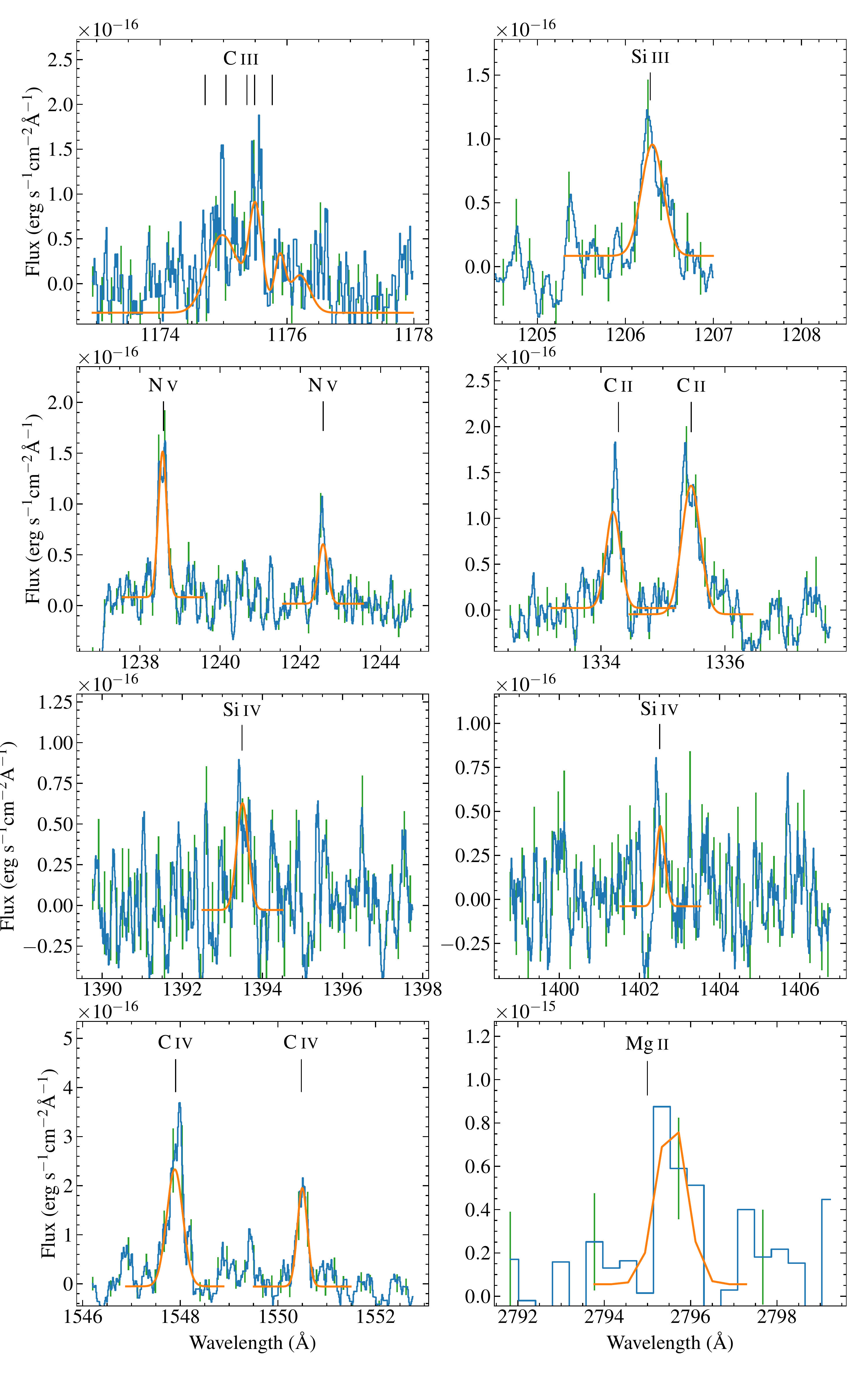}
    \caption{Detected emission lines in COS spectra of TRAPPIST-1 (blue) and the fit used to measure the integrated fluxes (orange). Example error bars on the flux values are shown in green. The data were smoothed with a 10-point boxcar for clarity, with the exception of the region around the \ion{Mg}{2} lines. Line positions were taken from NIST and shifted by the $-56.3$\,km\,s$^{-1}$ radial velocity of TRAPPIST-1 \citep{reiners+basri09-1}.}
    \label{fig:t1lines}
\end{figure}

\begin{table}
    \centering
    \begin{tabular}{llc}
Species & $\lambda_{rest}$ (\AA) & Flux ($10^{-18}$\,erg\,s$^{-1}$\,cm$^{-2}$) \\ \hline 
\ion{C}{3}$^m$ & 1174 & $101\pm27$ \\ 
\ion{Si}{3} & 1206.51 & $25.0\pm6.3$ \\
\ion{H}{1} (Ly\,$\alpha$) & 1215,67 & $1.40^{+0.60}_{-0.36}\times10^4$ \\ 
\ion{N}{5} & 1238.821 & $36\pm6.1$ \\ 
\ion{N}{5} & 1242.804 & $16\pm10.0$ \\ 
\ion{Si}{2} & 1264.73 & $\leq1.1$ \\ 
\ion{Si}{2} & 1298.95 & $4.4\pm3.5$ \\ 
\ion{Si}{2} & 1309.28 & $5.7\pm2.6$ \\ 
\ion{C}{2} & 1334.532 & $30\pm20$ \\ 
\ion{C}{2} & 1335.708 & $55\pm19$ \\ 
\ion{Cl}{1} & 1351.657 & $\leq3.1$ \\ 
$[$\ion{O}{1}$]$ & 1355.598 & $\leq7.6$ \\ 
\ion{O}{5} & 1371.292 & $\leq7.1$ \\ 
\ion{Fe}{2} & 1391.08 & $\leq2.5$ \\ 
\ion{Si}{4} & 1393.755 & $25\pm19$ \\ 
$[$\ion{O}{4}$]$ & 1401.156 & $\leq1.8$ \\ 
\ion{Si}{4} & 1402.77 & $14.0\pm7.6$ \\ 
\ion{Si}{2} & 1526.71 & $\leq3.2$ \\ 
\ion{Si}{2} & 1533.43 & $\leq3.2$ \\ 
\ion{C}{4} & 1548.195 & $100\pm46$ \\ 
\ion{C}{4} & 1550.77 & $53\pm17$ \\ 
\ion{C}{1} & 1561.0 & $\leq6.2$ \\ 
\ion{He}{2}$^m$ & 1640 & $\leq10.1$ \\ 
$[$\ion{O}{3}$]$ & 1666.153 & $\leq21.5$ \\ 
\ion{Al}{2} & 1670.787 & $\leq79$ \\ 
\ion{Al}{1} & 1766.39 & $\leq44$ \\ 
\ion{Mg}{2} & 2795.523 & $620\pm29$ \\ 
\ion{Mg}{2} & 2802.697 & $\leq340$ \\ 
\ion{Ca}{2} & 3933.6614  & $170\pm50$ \\ 
\ion{Ca}{2} & 3968.4673 & $\leq350^*$ \\

    \end{tabular}
    \caption{Integrated fluxes for ultraviolet emission line lists compiled by \citet{linsky17-1} and \citet{peacocketal19-1}. Lines at wavelengths not covered by our observations were omitted. $^*$Detected, but blended with H\,$\epsilon$. $^m$ Multiplet.}
    \label{tab:t1_line_fluxes}
\end{table}

\section{Emission line flux measurements}
\label{sec:em_lines}
Producing the DEM and semi-empirical models discussed below required identifying and measuring the fluxes of emission lines in the COS spectra. Table \ref{tab:t1_line_fluxes} provides fluxes for all of the lines in the lists compiled by \citet{linsky17-1} and \citet{peacocketal19-1} that are covered by our spectra. Where lines from the list were detected, we fit them using a Gaussian profile combined with a linear fit to the surrounding continuum to account for incorrect background subtraction  (Figure \ref{fig:t1lines}). The Gaussian profile does not always exactly recreate the apparent shape of the line profiles (for example the \ion{C}{2} lines), but, given the low S/N ratio, we cannot be confident that using more complex profiles would not lead to overfitting. We therefore do not claim to recover the shape of the line profile, but provide a reasonable measurement of the integrated flux. The flux is given as the integral of the Gaussian adjusted by the y-value of the linear fit, along with the propagated statistical error of the fit. Where the fit was unsuccessful we computed a 3\,$\sigma$ upper limit by treating the line as a Gaussian with a width fixed to the average of nearby lines and an amplitude equal to three times the error on the linear fit, adjusted by the linear fit as before. The \ion{Al}{2}\,1670.787\,\AA\ is formally detected by this method, but as the noise in the spectrum surrounding the line is of comparable amplitude we give its flux as an upper limit. Due to the complexity of the line profile of the \ion{C}{3}\,1176\,\AA\ multiplet, the flux was measured as the integral of the flux between 1174.5--1176.5\AA\, with the uncertainty estimated as the sum of the RMS of two adjacent 1\,\AA\ wide regions.

\section{Differential Emission Measure}
\label{sec:dem}

\begin{figure}
    \centering
    \includegraphics[width=0.5\columnwidth]{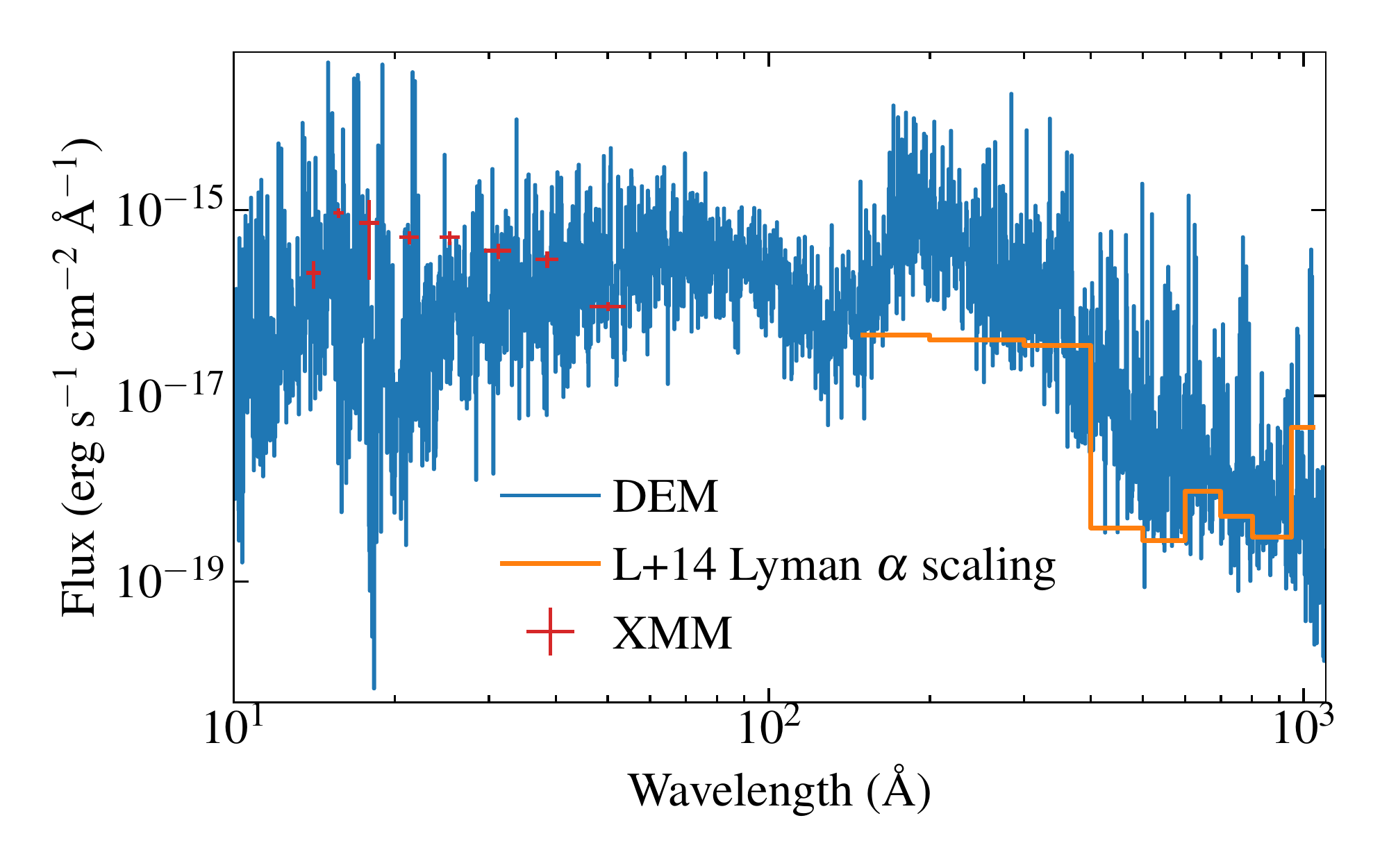}
    \caption{Comparison of the DEM model with the EUV/\lya\ relationships of \cite{linskyetal14-1}, and the \textit{XMM} spectrum.}
    \label{fig:dem}
\end{figure}

The differential emission measure (DEM) describes the amount of emitting plasma as a function of temperature for an optically thin plasma in coronal equilibrium \citep{warrenetal98-1, loudenetal17-1}, allowing us to predict the fluxes of emission lines formed in the coronal atmosphere if we have a functional form to describe the DEM. The DEM combined with the emissivity of a particular emission line determines the emission flux in that line emitted by a star. By measuring the flux of FUV emission lines formed at $\sim 10^5$ K and the X-ray flux formed at $ \gtrsim 10^6$ K, we can constrain the DEM by assuming it is well-described by a smooth low-order polynomial across the temperature range relevant to the chromosphere, transition region, and corona. This operates under the same principle as the \citet{linskyetal14-1} empirical scaling relations, but is tailored more specifically to the star by using more lines to resolve the structure of the star's upper atmosphere at a finer temperature resolution. Combining the differential emission measure with atomic data from \texttt{CHIANTI v8.0} \citep{dereetal97-1, delzannaetal15-1}, we estimate the EUV spectrum of TRAPPIST-1 between 10 to 912 $\textrm{\AA}$, incorporating the errors in fitting the differential emission measure forward to the predicted EUV spectrum. Full detailed of our DEM prescription are provided in \citet{duvvurietal21-1}.

\section{Semi-empirical model}
\label{sec:semp}

\begin{figure}
    \centering
    \includegraphics[width=\columnwidth]{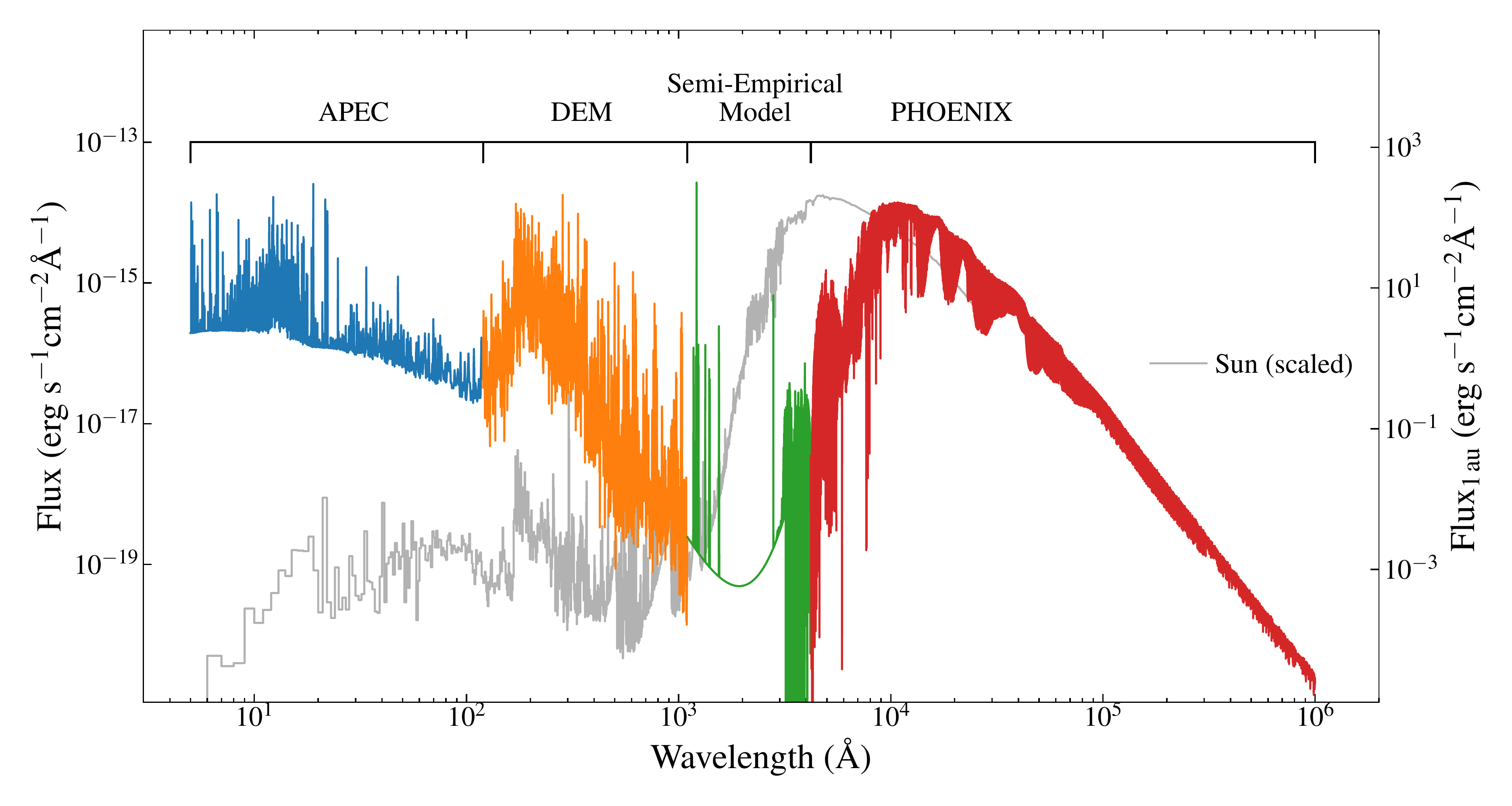}
    \caption{Semi-empirical model spectrum of TRAPPIST-1. The four different models components used are labeled above the spectrum, and the PHOENIX model has been rebinned to 1\,\AA\ for clarity. The spectrum is compared with the quiet Solar spectrum \citet{woodsetal09-1}, scaled to have the same blackbody photospheric flux as TRAPPIST-1. The left axis shows the flux received from TRAPPIST-1 at Earth, whereas the right shows the flux received at at the 1\,au equivalent distance, i.e. the spectrum of TRAPPIST-1 at a distance receiving the same photospheric flux as at 1\,au from the Sun}
    \label{fig:semp}
\end{figure}{}

Given the unavoidable low signal-to-noise of ultraviolet observations of late M-dwarfs in general and TRAPPIST-1 in particular, the Mega-MUSCLES data products will also include semi-empirical models constructed from the observed spectra, which we recommend as inputs for model planetary atmosphere studies as they avoid potentially biasing the models with the large amount of noise in the observed spectra. The semi-empirical model for TRAPPIST-1 is shown in Figure \ref{fig:semp}. Four models are used, including the APEC, DEM and PHOENIX models already discussed. The fourth model replaces the \textit{HST} spectra covering 1100--4200\,\AA. The model is constructed by first fitting a polynomial to the DEM model and the blue end ($\lambda < 4200$) of the STIS G430L spectrum to create a baseline flux. Note that this does not represent just the chromospheric/coronal continuum emission, but rather the combination of a forest of weak emission lines plus any continuum emission in that wavelength range. We then added the reconstructed Lyman $\alpha$ line and the fits to the emission lines shown in Figure \ref{fig:t1lines}, as well as the PHOENIX spectrum at those wavelengths, to produce the final spectrum section. 

\begin{figure}
    \centering
    \includegraphics[width=0.5\columnwidth]{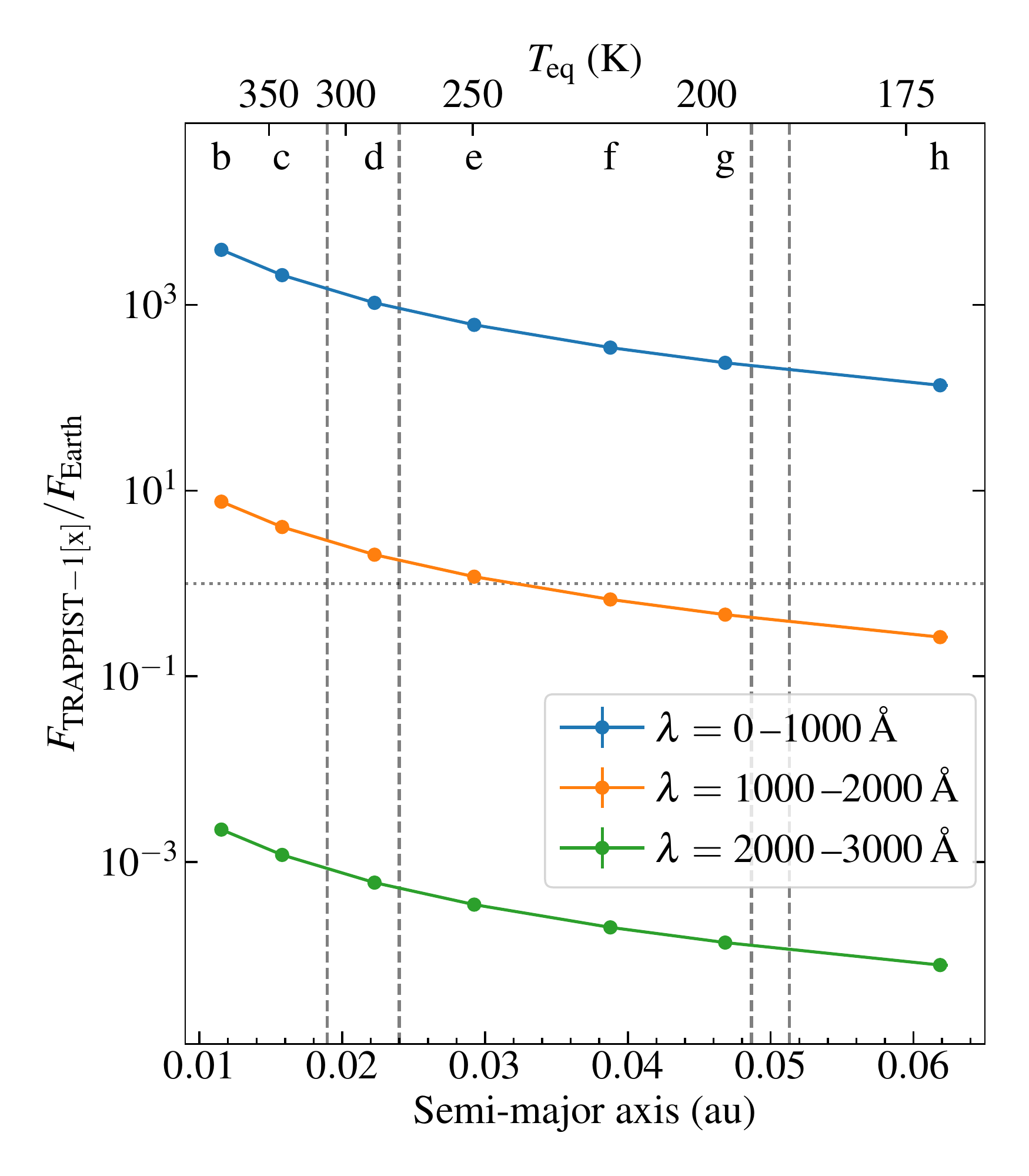}
    \caption{Fluxes at different wavebands experienced by the TRAPPIST-1 planets compared with the Earth. [x] in the y-axis label refers to planet b, c, etc.. The dashed lines show empirical (including TRAPPIST-1\,d) and conservative  (excluding TRAPPIST-1\,d) edges of the temperate zone for 1\,M$_{\bigoplus}$ planets as modelled by \citet{kopparapuetal14-1}. The equilibrium temperature as a function of distance from the star is shown on the top axis.}
    \label{fig:planets}
\end{figure}

Figure \ref{fig:semp} compares our final semi-empirical model SED with the quiet Solar spectrum \citep{woodsetal09-1}. The Solar spectrum is scaled by the ratio of the blackbody luminosities of the photospheres of the two stars, such that the photospheric flux  for each spectrum is the same as the irradiation Earth receives. The comparison clearly demonstrates the relative difference in high-energy flux between TRAPPIST-1 and the Sun, implying that any planet receiving the same photospheric flux from TRAPPIST-1 as (for example) the Earth receives from the Sun, is experiencing high-energy flux levels that are several orders of magnitude higher (Figure \ref{fig:planets}), even when the star is not flaring. 

\begin{figure}
    \centering
    \includegraphics[width=\columnwidth]{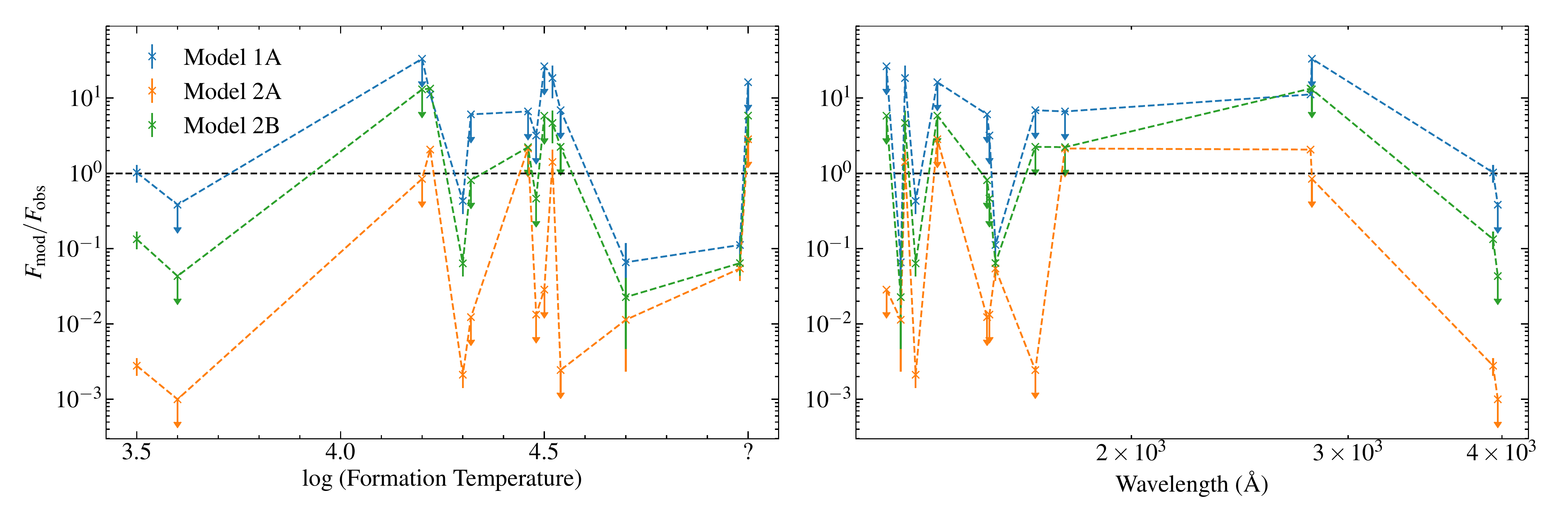}
    \caption{Comparison of the integrated flux of detected emission lines with predicted fluxes given in Table 5 of \citet{peacocketal19-1}  as  function of formation temperature (left) and wavelength (right). We were unable to find a measured formation temperature for \ion{Fe}{2}\,1391\,\AA.}
    \label{fig:modcomp}
\end{figure}

\section{Comparison with model spectra}
\label{sec:modcomp}

\citet{peacocketal19-1} extended the PHOENIX stellar atmosphere code into the ultraviolet by adding contributions from the chromosphere and transition regions, providing model SEDs of the TRAPPIST-1 ultraviolet regions against which we can compare our observations. The three PHOENIX models were calibrated to the \citet{bourrieretal17-1} \lya\ flux (model 1A) and distance-adjusted GALEX photometry of stars with similar spectral types (models 2A and 2B).

Figure \ref{fig:modcomp} compares the predicted line strengths given in Table 5 of \citet{peacocketal19-1} with the line fluxes measured from our COS data (Table \ref{tab:t1_line_fluxes}) as a function of formation temperature and wavelength.
For all three models, the agreement between the predicted and measured line fluxes is in general poor. The \lya-scaled model 1A accurately predicts the measured fluxes of the \ion{C}{2} and \ion{Ca}{2} lines, but predicts multiple lines at values $\sim 10$ times higher than the upper limits placed on their fluxes here. We find no trend between the accuracy of the predicted fluxes and either formation temperature or wavelength.

\section{Swift photometry}
\label{sec:swift}
As this paper was under review, \cite{beckeretal20-1} presented a deep Swift \texttt{uvm2} observation of TRAPPIST-1, finding a flux of $8.4\pm1.4\times10^{-15}$\,erg\,s$^{-1}$\,cm$^{-2}$ around 1900\,\AA. As a large fraction of the \texttt{uvm2} waveband is covered by the gap in the COS G230L detector, an exact comparison between the SED and the Swift measurement is challenging. Integrating the semi-empirical model over the \texttt{uvm2} waveband, we find a flux of $1.5\times10^{-15}$\,erg\,s$^{-1}$\,cm$^{-2}$, suggesting that the model may underpredict the flux in the NUV, although there is no way to further test this with the available data. We note that, although the \texttt{uvm2} peaks at $\approx1900$\,\AA\, the effective area is still $\approx10$\,percent of maximum around the \ion{Mg}{2}\,2800\,\AA\ lines. As these are the only NUV features detected in our COS spectroscopy, it is possible that they are contributing the majority of the flux detected by Swift. Similar photometric measurements of emission lines may be an avenue to study the ultraviolet SEDs of low-mass stars too faint to observe spectroscopically.

\section{High-Level Science Products}
The TRAPPIST-1 SED will be made available at or before publication of this paper on the MUSCLES Treasury Survey Page at the Mikulski Archive for Space Telescopes (MAST): \url{https://archive.stsci.edu/prepds/muscles/}, \url{https://doi.org/doi:10.17909/T9DG6F}. Available products will include the standard data products provided by the MUSCLES survey (i.e., the SED at native and 1\,\AA\ resolutions along with the component observations and models), along with the new Semi-empirical model SED at native and 1\,\AA\ resolutions. The remainder of the Mega-MUSCLES targets, listed at \url{http://cos.colorado.edu/~kevinf/muscles.html}, will be added to the database in the coming months.\footnote{ArXiv copy note: Due the first author living in Texas in February 2021, the HLSP products will not be available when this paper goes on arXiv. For now, the semi-empirical SEDs can be found here: \url{https://github.com/davidjwilson/Trappist-1_MM}.}

\section{Conclusion}

We have constructed a panchromatic SED of the M8 star TRAPPIST-1, the first data product from the Mega-MUSCLES survey. 

TRAPPIST-1 is the faintest target in the Mega-MUSCLES survey, and the SED presented here both represents the state-of-the-art for observation of the high-energy flux of low-mass stars and demonstrates the limits of our current observing facilities. Obtaining the ultraviolet spectroscopy pushed the capabilities of COS to their limits, with many expected emission lines remaining below the noise limit. The EUV spectrum cannot be observed with any currently operating facility. Improving on these observations, which is desirable given the continued importance of low-mass stars for exoplanet science, will require the launch of large-aperture space telescopes with ultraviolet capabilities and a dedicated EUV observatory \citep{youngbloodetal19-1}. 


\acknowledgments
\noindent We thank S. Peacock for providing the PHOENIX EUV models and E. Gonzales for providing the optical spectra and photometry. Based on observations made with the NASA/ESA Hubble Space Telescope, obtained from the Data Archive at the Space Telescope Science Institute, which is operated by the Association of Universities for Research in Astronomy, Inc., under NASA contract NAS 5-26555. These observations are associated with program \# 15071. Support for program \#15071 was provided by NASA through a grant from the Space Telescope Science Institute, which is operated by the Association of Universities for Research in Astronomy, Inc., under NASA contract NAS 5-26555. All of the \textit{HST} data presented in this paper were obtained from the Mikulski Archive for Space Telescopes (MAST). AY acknowledges support by an appointment to the NASA Postdoctoral Program at Goddard Space Flight Center, administered by USRA through a contract with NASA. PCS acknowledges support by DLR under grant 50 OR 1901.
 
%

\vspace{5mm}
\facilities{\textit{HST} (STIS and COS), \textit{XMM-Newton}}

\defcitealias{astropy18-1}{Astropy Collaboration, 2018}
\software{astropy \citepalias{astropy18-1}, xpec \citep{arnaud96-1}, stistools\footnote{\url{https://stistools.readthedocs.io/en/latest/}}, scipy \citep{virtanenetal20-1}, numpy \citep{harrisetal20-1}, matplotlib \citep{hunter07-1}}



\appendix




\bibliographystyle{aasjournal}
\bibliography{aabib}



\end{document}